"Reconsidering Experiments"

Lydia Patton

Abstract

Experiments may not reveal their full import at the time that they are performed. The scientists who perform them usually are testing a specific hypothesis, and quite often have specific expectations limiting the possible inferences that can be drawn from the experiment. Nonetheless, as Hacking has said, "experiments have lives of their own". Those lives do not end with the initial report of the results and consequences of the experiment. Going back and re-thinking the consequences of the experiment in a new context, theoretical or empirical, has great merit as a strategy for investigation and for scientific problem analysis. I apply this analysis to the interplay between Fizeau's classic optical experiments and the building of special relativity. Einstein's understanding of the problems facing classical electrodynamics and optics in part was informed by Fizeau's 1851 experiments. However, between 1851 and 1905, Fizeau's experiments were duplicated and reinterpreted by a succession of scientists, including Hertz, Lorentz, and Michelson. Einstein's analysis of the consequences of the experiments is tied closely to this theoretical and experimental tradition. However, Einstein's own inferences from the experiments necessarily differ greatly from the inferences drawn by others in that tradition.

1. *Experiment, theory, and matter-ether interaction*
    a. *Stellar aberration and Fizeau's experiments*
    b. *Lorentz's reconsideration of the Fizeau experiments*
    c. *Hertz's response to the early Michelson experiments*
2. *Einstein's 1905 paper*
3. *Reconsidering experiments*

Experiments have lives of their own, in Hacking's well-known formulation. Those lives do not end with the experiments themselves, but include how researchers take up the results of the experiments, in formulating and pursuing theories, problems, and research programs. (Franklin 1986 and 1993) and (Hacking 1983), among others, emphasize the role of experiment in the building and pursuit of scientific theories, following Peirce's and Laudan's earlier analyses. I focus on how successive reconsideration of key experiments within a given research tradition can reveal substantive information about the problems scientists see themselves as needing to solve, and about the structure of possible new theories and research programs.

I investigate the relationship between the experimental and theoretical tradition in electrodynamics and optics in the $19^{th}$ century and Einstein's abandonment of the luminiferous ether as "superfluous" in his 1905 paper on special relativity. Einstein's strategy here was partly a very clever avoidance of blind alleys into which $19^{th}$ century researchers had been led; in particular, the blind alley of including matter-ether interaction as a variable in electrodynamics and optics. I will trace a tradition of inferences about matter-ether interaction drawn from the Fizeau experiments, a tradition of which Einstein was aware. Einstein's knowledge of this tradition informed his decision to build a theory with certain distinctive features. Einstein's removal of the ether required him to depart from the series of previous conclusions drawn from the Fizeau experiments in particular, conclusions which blocked the principle of relativity.

I conclude that the role of re-examining experiments sometimes is to show that a given problem, like the problem of matter-ether interaction, has hitherto unknown facets or even solutions. This fact can be of considerable interest, independently of the other material conclusions that can be drawn from the experiment.

1. *Experiment, theory, and matter-ether interaction*

The influence of experiment and of 19th century ether theories on the building of special relativity is difficult to assess. Einstein gives few citations in his 1905 paper on electrodynamics, and few letters, notes, drafts, or journal entries from the period survive.[i] In his sixties, Einstein writes that, during his student years, he read works by Hertz, Maxwell, Kirchoff, and Helmholtz (Einstein 1979/1946, 15). Einstein presents his argument against the luminiferous ether in 1905 as partly inspired by Mach's Humean arguments that the ether and atoms should be treated as speculative and not as well-founded empirically. In the same reminiscences, though, Einstein argues that "exterior" criteria, testing the fit between theory and observed phenomena, are crucial and primary (Einstein 1979/1946, 20ff).

(Holton 1969) cites a first-hand report by Shankland, according to which Einstein said he could not remember whether he was aware of Albert Michelson's ether drift experiments when he composed his 1905 electrodynamics paper (p. 154). In that paper, Einstein does mention "unsuccessful attempts to discover any motion of the earth relatively to the 'light medium,'" but these could have been Bradley's and Airy's stellar aberration experiments (Einstein 1952/1905, 37). Shankland reports from those conversations that Einstein said he was aware of the stellar aberration experiments, and of Hippolyte Fizeau's experiments from the 1850s, testing the effect of traveling through refractive media, such as water, on the velocity of light—and that these two sets of experiments were "enough".[ii]

But to say that Einstein was aware of Fizeau's experiments and stellar aberration is not to say that these were an *experimentum crucis* that decided the question in favor of special relativity, or to say that the experiments pointed unambiguously and immediately to the theory of special relativity.[iii] To say so is to skip two significant steps. First, as emphasized recently by (Stachel 2005), 19th century scientists did not see the Fizeau experiments as a crucial test ruling out ether theories. Fizeau himself intended the experiments to help determine the correct approach to a theory of matter-ether interaction, and the approaches he considered are incompatible with the principle of relativity. Only in 1895 did Hendrik Lorentz show that the Fizeau experiments could be reinterpreted without matter-ether interaction, and even Lorentz retains the immobile ether as a reference frame.

The second step, then, is Einstein's reinterpretation of the Fizeau experiments in the light of Lorentz's and Hertz's subsequent theories. In §6 of the 1905 paper, and in his later lecture on the ether and relativity (Einstein 2007/1920), Einstein works from a very different interpretation of the Fizeau experiments from Fizeau's own. This distinct analysis points the way for Einstein to narrow down possible strategies for building his theory of electrodynamics and of optics. In doing so, he considers carefully the way that Lorentz's and Hertz's theories incorporate the ether into their explanations. I will consider these two steps in turn.

a. *Stellar aberration and Fizeau's experiments*

Experiments on the angle of aberration of light from a star reaching a telescope on the moving earth were a prominent stimulus for Fizeau's experiments.[iv] An early such experiment was performed in 1727, by James Bradley. Norton points out that Bradley's stellar aberration results were explicable either on the theory that light is emitted as a stream of particles from the star (the emission theory), or that light is a wave propagating in the luminiferous ether (the wave theory) (Norton, forthcoming, §4.5). It was worth an experimenter's while to attempt to find a way to decide between the emission and wave theories. In 1871, George B. Airy repeated the experiment, with a different experimental design. Airy filled the telescope with water, and then compared the results of the water-filled telescope with those of the ordinary telescope. (Holton 1969) sums up the results:

> On the model of light as wave propagation through an ether, the aberration angle was expected to be larger when the observing telescope was filled with water, but on experiment the angle was found to be the same. Augustin Fresnel therefore proposed that the ether is partly carried or dragged along in the motion of a medium (such as water) having a refractive index larger than 1 (p. 136).

The equation measuring the influence of the medium on the velocity of light became known as the "Fresnel drag coefficient".[v] Fresnel's drag coefficient could not be a direct measure of ether drag, since that is an unobservable. The drag coefficient measures only the velocity of light proportional to the index of refraction of the medium in which the light is moving.

Any good ether theory needed, then, to be able to relate the velocity of light, relative to the index of refraction of the medium through which the light travels, to a theoretical explanation in terms of ether drag. As Fizeau sums it up in an 1851 paper reporting his experimental results,

> Several theories have been proposed to account for the phenomena of light aberration in the wave system. First Fresnel, and more recently Monsieurs Doppler, Stokes, Challis and several others, have published important works on the subject; but it does not appear that any of the proposed theories has received the complete assent of physicists. In the absence of secure notions of the properties of the luminiferous ether and of its relations with ponderable matter, one has had to make hypotheses (Fizeau 1851, 385).

Fizeau's self-reported intention in constructing his experiment was thus to choose an appropriate *theory* to explain light aberration. The drag coefficient was well supported already by Arago's and Fresnel's results. But the drag coefficient itself does not explain light aberration, nor did it lend itself easily to constructing a theory. Fizeau shrewdly identifies the problem: there was no good theory of the interaction between the light ether and matter. Without such a theory, explanations of light aberration in terms of ether drag beg the crucial question: *how* does the ether participate in the behaviour of light with respect to the medium through which it travels?

As Fizeau sees it, then, a suitable theory of matter-ether interaction had not been found, and Fizeau's intention is for his experiment to help to decide among various hypotheses that could explain this interaction. As Fizeau conceives of them, these hypotheses are:

> Either the ether is adherent and as if fixed to the molecules of the body, and consequently shares the movements that can be attributed to this body;
>
> Or in fact the ether is free and independent, and is not dragged by the body in its movements;
>
> Or, finally, by a third hypothesis, which shares elements of the other two, only part of the ether would be free, and the other part would be fixed to the molecules of the body and it alone would participate in the movements of that body (Fizeau 1851, 349-350).

The third hypothesis is Fresnel's.

> Fizeau then analyzes each hypothesis, and observes that
>
> for each of these hypotheses, the influence of motion on the speed of light will be different:
>
> If the ether is entrained with the body in its motion, the speed of light will be increased by the entire speed of the body, the ray being supposed to be directed towards the motion;
>
> If the ether is supposed free, the speed of light will not be altered at all;
>
> If only a part of the ether is entrained, the speed of light will be increased [when the ray is traveling with the current], but by only a fraction of the speed of the body and not by the totality of it as with the first hypothesis. This consequence is not as evident as the two preceding ones, but Fresnel has shown that it can be supported by very probable mechanical considerations (Fizeau 1851, 386-387).

The experiment Fizeau devises to attempt to decide between the hypotheses is deceptively simple.

> The mode of observation […] consists in producing interference fringes with two rays of light, after passing through two parallel tubes, in which air or water flow with a great speed and in opposite directions (Fizeau 1851, 387-388).
>
> After the light rays go through the tubes of water or air, they interfere a little in a window that they pass through, and it is there that one observes the fringes, by means of an eyepiece with divisions (*Ibid*. 389).

(Mascart 1893, 101) provides this schematic diagram of Fizeau's experiment:[vi]

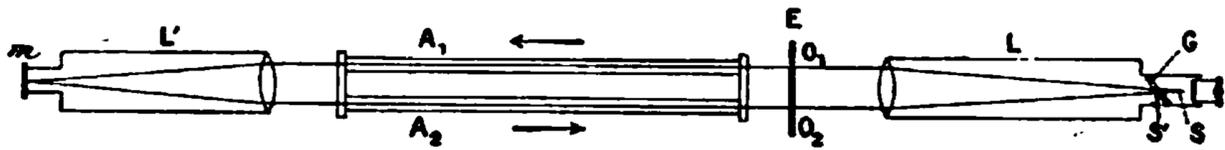

The observer is at the far right. L and L′ are converging lenses, and *m* on the far left is a mirror. E is a screen, and $O_1$ and $O_2$ are slits in the screen. G is the window or sheet of glass on which the "fringes" are observed. Fringes are the visible sign of the location of the light rays at a certain time. S′ and S are the locations of the fringes. $A_1$ and $A_2$ are tubes of water, five millimeters in diameter and nine millimeters apart. The water in $A_1$ flows in the opposite direction from the water in $A_2$.

Two rays of sunlight are sent through both tubes, in succession, beginning at the same time. Each is sent through one tube, converging onto the mirror, reflecting back through the other tube, and then converging again, to produce a fringe on the window. One ray travels with the current, and another against the current. If the light travels at the same speed no matter in which direction the medium is moving, then the fringes should coincide perfectly in the "window", given that they take a path of the same length through the same medium. If they do not coincide perfectly, then "displacement of the fringes" will be observed, that is, an area in which one fringe is seen before the other fringe, because the light is moving faster or slower. This appears to the observer as the fringes moving to the left or to the right for a few instants.

Fizeau reports the following results:

> *[T]he motion of the air does not produce any sensible displacement of the fringes.* […] For water, there was an evident displacement. […] The fringes moved to the right, when the water moves away from the observer in the tube to his right, and towards the observer in the tube to his left. The fringes moved to the left, when the direction of current in each tube was in the opposite direction (Fizeau 1851, 390-391, original emphasis).

From a backward-looking relativistic perspective, this way of putting the observations is suggestive. The displacement of the fringes is symmetrical for both cases, in whichever direction the water is moving with respect to the observer. From this perspective, there is nothing in the observations to distinguish the two cases from each other except relative motion.

The case is very different when we turn to Fizeau's conclusions from the experiment, about which theoretical hypothesis to adopt. Fizeau argues that the displacement observed is consistent with the hypothesis that the ether is partly "dragged" with the current. But adopting the hypothesis of partial or full ether drag requires that the ether's motion depends on the motion of the material body that drags it, the water in this case. That requires some way to distinguish whether the ether is moving with respect to the water, or the water with respect to the ether – that is, it requires distinguishing inertial frames from each other.

Fizeau adopts the hypothesis of partial ether drag, with evident reluctance. He argues that the fact that any displacement is observed at all rules out the second hypothesis of free ether (*Ibid*., 393). As Fizeau reasons above, if the ether is free, the velocity of light will not be altered at all by the medium. Fizeau argues that the actual displacement values are not great enough to confirm the first hypothesis, of the fully dragged ether (*Ibid*., 395). Fizeau is forced to concede that "the third hypothesis, that which is due to Fresnel, leads to a value of displacement very little different from the result of the observation" (*Ibid*., 396). Fizeau concludes that Fresnel's hypothesis of partial ether drag "appears" to give a theoretical explanation for the experiments.

Fizeau has scruples about this result, however. As Stachel observes, Fizeau concludes his paper with the remark that while Fresnel's drag coefficient is well supported by his experiments,

> Fresnel's conception [of partial ether drag] would appear so extraordinary, and in
> several respects so difficult to accept, that one would require still more proofs and
> a deepened examination by mathematical physicists [*géomètres*], before accepting it as the expression of the way things really are (Fizeau 1851, 404, translation from Stachel 2005, 6).

b.  *Lorentz's reconsideration of Fizeau's experiments*

As Stachel (2005) traces in detail, Fizeau's experiment was not accepted at the time as a rigorous test of theories of ether drag. The responses to Fizeau's experiment in the succeeding years, by Mascart, Poincaré, and Veltmann, for instance, recognize Fizeau's result as a significant empirical advance, but not as giving a clear signpost on the path to an adequate theoretical explanation (pp. 6-7). Fizeau frames his experiment in 1851 as a test of the explanatory hypotheses about the interaction between matter and ether: is the ether free of interaction with matter in motion, is it partially dragged by matter in motion, or is it fully dragged? But Fizeau never considers the hypothesis that the entire effect could be explained without appealing to matter-ether interaction at all.

A signal advance on this score was made in 1895 by Hendrik Lorentz, in his well-known "Attempt at a theory of electrical and optical phenomena in moving bodies", or, as it is usually called, the *Versuch*. Lorentz shows that Fresnel's "drag" coefficient can be interpreted as exclusively a refraction coefficient (Lorentz 1895, 82ff.). Lorentz proves that the effect observed by Fizeau can be explained—and predicted—by the reflection and refraction of light waves by the medium through which they pass.[vii]

In fact, then, the first reconsideration of the experiment in this narrative takes place in the *Versuch*. Fizeau's intention with his original experiment was to test three hypotheses about matter-ether interaction, but Fizeau's write-up of his experiment does not consider an alternative hypothesis: that there is no matter-ether interaction involved, rather, some other cause is responsible for the observed effect. Lorentz considers this hypothesis, and concludes that it is correct: the reflection and refraction of light waves, not matter-ether interaction, is responsible for the effects Fizeau observed.

In his optical and electrodynamical theories, Lorentz does not abandon the ether. He retains the "still ether" as a reference frame. In his (1895), Lorentz distinguishes between motion of a light ray with respect to matter and with respect to the ether, for instance (p. 97).

However, as Einstein puts it in his Kyoto lecture (Einstein 2007/1920),

> As to the mechanical nature of the Lorentzian ether, it may be said of it, in a somewhat playful spirit, that immobility is the only mechanical property of which it has not been deprived by H.A. Lorentz (p. 615).

Einstein sees Lorentz's removal of most mechanical properties from the ether, such as Fresnel's ether density, as a significant step toward the correct account. However, Lorentz's "still ether" is an absolute reference frame that distinguishes inertial frames from each other. As such, a picture of electrodynamics and optics that includes the still ether blocks the principle of relativity.

*c.     Hertz's response to the early Michelson experiments*

Lorentz's influence on Einstein's 1905 paper, and on Einstein's subsequent elaborations of the theory, is well established. However, Einstein's famous assertion in the beginning of the 1905 paper, that he will not "assign a velocity-vector to a point of the empty space in which electromagnetic processes take place", has another source: the electrodynamic theory of Heinrich Hertz (Einstein 1952/1905, 38). There is documentary evidence from the period in which Einstein was constructing special relativity that Einstein was reading Hertz's work on electrostatics and electrodynamics. In a letter to Mileva Mariç of 1899, Einstein writes:

> I've returned the Helmholtz volume and am once again studying Hertz's *Investigations into the Propagation of Electric Force* very carefully. […] I am becoming more and more convinced that the electrodynamics of moving bodies, as it is currently presented, does not correspond to reality and instead can be presented more simply. The introduction of the name 'ether' into electrical theories has led to the idea of a medium of whose movement one may be able to speak, yet in my opinion without being able to link a physical sense to this expression (Einstein to Mariç 1899, cited in Einstein 1987, 226).

After Hertz's famous experiments on radio waves in the late 1880s, Hertz, with many other German scientists, leaned toward Maxwell's field theory as the simplest and most elegant explanation of the observed phenomena.[viii]

However, in his (1865), Maxwell incorporated states of tension and elasticity of the ether, which he saw on analogy with an incompressible fluid, into his model of electrodynamic action.[ix] As a report of a lecture by Einstein sums up the situation,

> For Maxwell himself the ether indeed still had properties which were purely mechanical […] But neither Maxwell nor his followers succeeded in elaborating a mechanical model for the ether which might furnish a satisfactory mechanical interpretation of Maxwell's laws of the electromagnetic field. The laws were clear and simple, the mechanical interpretations clumsy and contradictory (Einstein 2007/1920, 614).

In *Investigations into the Propagation of Electric Force*, Hertz tries to assign a more restricted role to the ether in his own theory than it is given in Maxwell's, based on the negative results of experiments attempting to establish the motion of the ether relative to matter.

Hertz's electrostatic theory, a theory of charges and potentials, assigns a "single directed magnitude," a vector, to each point of space, to describe the "electric and magnetic state" of "the medium which fills the space at that point" (Hertz 1900/1890b, 241). In the case of actions that take place in the ether and no other medium, the ether is assigned a vector nonetheless. The significant point is that charges and potentials are taken to depend on conditions of the ether. Hertz proceeds to identify a difficulty with applying his approach to electrodynamics, the description of currents and changing fields:

> whenever in ordinary speech we speak of bodies in motion, we have in mind the motion of ponderable matter alone. According to our view, however, the disturbances of the ether, which simultaneously arise, cannot be without effect; and of these we have no knowledge. This is equivalent to saying that the question here raised cannot at present be treated at all without introducing arbitrary assumptions as to the motion of the ether. Furthermore, the few existing indications as to the nature of the motion of the ether lead us to suppose that the question above raised [whether Hertz's approach in electrostatics can be extended to electrodynamics] is strictly to be answered in the negative, for it appears to follow from such indications as we have, that even in the interior of tangible matter the ether moves independently of it; indeed, this view can scarcely be avoided in view of the fact that we cannot remove the ether from any closed space. If now we wish to adapt our theory to this view, we have to regard the electromagnetic conditions of the ether and of the tangible matter at every point in space as being in a certain sense independent of each other (Hertz 1900/1890b, 241ff.).

Hertz's initial, preferred approach was to consider the charge of matter in a neighborhood to depend on the electromagnetic conditions of the ether in that neighborhood. This allowed him to exclude Helmholtz's action at a distance. In electrostatics, there was no bar to positing this dependence, although there was no particular evidence for it either. But in electrodynamics, there was a barrier: the Fizeau and Michelson experiments.[x]

Michelson was engaged in duplicating Fizeau's experiment when he invented the interferometer in his early experiments in Berlin, in 1881. Michelson and Morley re-cast the conclusions one can draw from Fizeau's experiment when they performed an expanded version of the experiment in 1885-1886. In these experiments, Michelson and Morley famously try to detect ether drift with respect to the earth's motion. However, the experiments were also intended to repeat the Fizeau experiments by investigating more closely, with the new, more precise interferometer, "the influence upon the velocity of light of the motion of the medium through which it passes" (Michelson and Morley 1886, 377). While Michelson, Morley, and Fizeau agree that the Fresnel drag coefficient fits the data, Fizeau argues that his results support hypothesis of a partially dragged ether. Michelson and Morley argue the experiments support a modified claim: "the luminiferous ether is entirely unaffected by the motion of the matter which it permeates" (*Ibid*, 386).

Michelson's and Morley's reading does not correspond directly to any of Fizeau's three hypotheses, but is rather a more general conclusion. It appears at first glance to correspond to Fizeau's second hypothesis, of the free ether. But Michelson and Morley argue that another interpretation is possible. They report Fresnel as arguing that the ether within a moving body is stationary except those parts coalesced around the moving particles of the body, which is Fizeau's third hypothesis. Michelson and Morley respond that if each particle of a body, with its "halo" of ether, is treated as a single body, then the hypothesis that the ether is not affected by the motion of the body is confirmed by their more precise results (*Ibid*, 379). If the ether is not separate from the particles of the body, then no variables are needed to describe the effect of the motion of matter on the ether.

Hertz continues his essay on electrodynamics by arguing that responding to experimental data requires removing any independent properties of the ether from electrodynamics, since experiments had failed to detect the motion of the ether relative to matter:[xi]

> But the state of the case is different if we explicitly content ourselves with representing electromagnetic phenomena in a narrower sense—up to the extent to which they have hitherto been satisfactorily investigated. We may assert that among the phenomena so embraced there is not one which requires the admission of a motion of the ether independently of ponderable matter within this latter; this follows at once from the fact that from this class of phenomena no hint has been obtained as to the magnitude of the relative displacement. At least this class of electric and magnetic phenomena must be compatible with the view that no such displacement occurs, but that the ether which is hypothetically assumed to exist in the interior of ponderable matter only moves with it [...] For the purpose of the present paper we adopt this view (Hertz 1900/1890b, 242).

Thus, Hertz solves the problem posed by experiment for electrodynamic theory by postulating the fully dragged electromagnetic ether, which is consistent with Michelson's and Morley's later reading of the consequences of their experiments.

*2.  Einstein's 1905 paper*

As Stein (1970) and Stachel (2005) show, the problem of detecting and describing ether-matter interaction was perhaps the most significant problem for ether theories throughout the 19th century.[xii] In the 1905 paper and in the 1920 address, Einstein works from a re-interpretation of the Fizeau experiments, and uses this re-interpretation to remove barriers to his argument for the principle of relativity. Fizeau concludes reluctantly that the ether is partially dragged by matter in motion, though he concedes that this hypothesis is implausible and tentative. Hertz infers from the experiments that the ether is not affected by the motion of matter and does not move independently of it. Nonetheless, Hertz retains an electrodynamic ether fully dragged by matter. Even Lorentz retains the "still ether" as a reference frame (Lorentz 1895, 97). Einstein argues that a re-casting of the experiments justifies the claim that the ether is "superfluous" (Einstein 1952/1905, 38) and "does not take part in the movements of bodies" (Einstein 2007/1920, 614). This seemingly minor re-casting of the consequences of the experiments has profound significance for the interpretation of special relativity, and of its relationship to the preceding experimental and theoretical tradition.

In the introductory remarks to §6 of the 1905 paper, Einstein observes:

> The introduction of a "luminiferous ether" will prove to be superfluous inasmuch as the view here to be developed will not require an "absolutely stationary space" provided with special properties, nor assign a velocity-vector to a point of the empty space in which electromagnetic processes take place (Einstein 1952/1905, 38).

Einstein's remark about "absolute stationary space" often is taken to be a response to Lorentz, and certainly it appears to be.[xiii] Lorentz focuses on whether it is possible to detect the motion of the ether relative to matter, and takes the null result of the ether experiments to confirm his theory of the stationary ether. Einstein takes this reasoning one step further: if the motion of the ether relative to matter can't be detected, why include it in the theory, as a postulated vector quantity, as a set of variables, or as a reference frame?

Einstein's first remark was no doubt directed at Lorentz. But what about the remark about assigning a velocity vector to empty space? In the *Versuch*, Lorentz argues that the divergence assigned to the "pure ether" is zero (Lorentz 1895, §5). By contrast, Hertz 1900/1890b begins with a proposal to assign velocity vectors to apparently empty space, which, Hertz postulates, is filled with the ether:

> We therefore assume that at every point a single definite velocity can be assigned to the medium which fills space, and we denote the components of this in the directions of $x$, $y$, $z$ by $\alpha$, $\beta$, $\gamma$. […] Wherever we find tangible matter in space we definitely deduce the values of $\alpha$, $\beta$, $\gamma$ from the motion of this. Wherever we do not find in the space any tangible matter, we may assign to $\alpha$, $\beta$, $\gamma$ any arbitrary value which is consistent with the given motions at the boundary of the empty space, and and of the same order of magnitude. We might, for example, give $\alpha$, $\beta$, $\gamma$ those values which would exist in the ether if it moved like any gas (Hertz 1900/1890b, 243).

Hertz removes Maxwell's variables describing the elasticity and density of the ether from the theory in Hertz 1900/1890b. But Hertz's 1890 theory requires postulating that the ether behaves like a gas, and assigning corresponding velocity vectors to empty space.

For Einstein, Hertz's theory wins the formal battle with Maxwell but loses the interpretive war.[xiv] Maxwell appeals to interaction between the ether and material bodies to explain electrodynamic action. Hertz argues that there is no evidence for action of the electromagnetic ether independent of matter. Nonetheless, for Hertz, the ether is the carrier of fields. As the report of Einstein's Kyoto lecture has it, in Hertz's theory,

> matter appears not only as the bearer of velocities, kinetic energy, and mechanical pressures, but also as the bearer of electromagnetic fields. Since such fields also occur in vacuo—i.e. in the ether—the ether also appears as bearer of electromagnetic fields. The ether appears indistinguishable in its functions from ordinary matter. Within matter it takes part in the motion of matter and in empty space it has everywhere a velocity; so that the ether has a definitely assigned

> velocity throughout the whole of space. There is no fundamental difference between Hertz's ether and ponderable matter (which in part subsists in the ether). The Hertz theory […] was […] at variance with the result of Fizeau's important experiment on the velocity of the propagation of light in moving fluids, and with other established experimental results (Einstein 2007/1920, 615).

The final sentence is not quite historically correct. As discussed above, Hertz's theory of the fully dragged ether was at least consistent with Michelson's and Morley's interpretation of their own results from 1885 and 1886.

But Hertz's theory is inconsistent with Einstein's own interpretation of the Fizeau result, which is that one is "obliged to infer" from Fizeau's experiment that "the luminiferous ether does not take part in the motions of bodies" (*Ibid*. 614). It is not correct that the Hertz theory was falsified by experiment, but it *is* correct that Hertz's inferences from experiment, along with Michelson's and Morley's own inferences from their experiments, were underdetermined by the experimental evidence. For Einstein, there is no evidence that the ether moves at all, and so there is no reason to assign it a velocity vector.

Further, if Hertz's and Michelson's postulate of the fully dragged ether were correct, and if Hertz's velocity vectors were assigned to empty space, then that would distinguish inertial frames from each other. The ether would have "a definitely assigned velocity throughout the whole of space". That would mean that the laws of electrodynamics and of optics are *not* invariant with respect to arbitrary transformations of inertial frames, which would undermine the principle of relativity.

Whether Einstein formulated the principle of relativity before or after his reconsideration of the consequences of the Fizeau experiment is an interesting and perhaps unanswerable question. However, even *once* Einstein had formulated the principle of relativity, two things were necessary for Einstein to be able to formulate an argument for the relativity of simultaneity as consistent with experiment:

1. Einstein had to be able to argue against Lorentz's still ether, which distinguishes inertial frames from each other.
2. Einstein had to argue that the ether does not "take part in" the motions of bodies, as do Hertz's fully dragged ether and Fizeau's partially dragged ether. If the ether is the carrier of fields, and the ether has "a definitely assigned velocity throughout the whole of space", then here again, the principle of relativity arguably would not hold. This possibility requires Einstein to reinterpret Hertz's and Fizeau's inferences from the Fizeau and Michelson experiments.

In the 1905 paper, Einstein demonstrates that there is a consistent and simple model of his own theory that eliminates matter-ether interaction altogether. Einstein shows that the single relation necessary to the representation of electrical action is a consistent representation of the force exerted on the charge carriers. He proves, using the Lorentz transform, that his model of that force is invariant with respect to arbitrary transformations of inertial frames. The "path and intensity" of the currents in the Maxwell-Hertz equations are represented by vectors. These vectors map displacements of points within systems of spatial coordinates. The principle of relativity postulates that the laws of electrodynamics and of optics are valid for arbitrary transformations of these systems of coordinates.

Einstein postulates the principle of relativity, the light principle, and the Maxwell-Hertz equations to capture the relations that Hertz and Maxwell capture using postulates about the motion and mechanical and dynamical behavior of the ether. Again, then, Einstein eliminates the relation of interaction between ether and matter altogether.

Einstein's removal of the ether from this theory is backed by a sound but little-appreciated re-interpretation of the Fizeau experiments, which differs significantly from the conclusions Fizeau, Michelson, Morley, and Hertz draw from the same experiments. The following are the conclusions drawn, beginning with Fizeau's third hypothesis, the one he accepts after the experiment:

> only part of the ether would be free, and the other part would be fixed to the molecules of the body and it alone would participate in the movements of that body (Fizeau 1851, 350).
>
> the luminiferous ether is entirely unaffected by the motion of the matter which it permeates (Michelson and Morley 1886, 386).
>
> the ether which is hypothetically assumed to exist in the interior of ponderable matter only moves with it (Hertz 1900/1890b, 242).
>
> The introduction of a 'luminiferous ether' will prove to be superfluous (Einstein 1952/1905, 38).
>
> the luminiferous ether does not take part in the movements of bodies (Einstein 2007/1920, 614).

Fizeau, Michelson and Morley, Lorentz, and Hertz focus on whether it is possible to detect the motion of the ether relative to matter. Einstein goes further: if the motion of the ether relative to matter cannot be detected, then the ether should not be considered to contribute to the motions of bodies or to move independently of or with material bodies—it is "superfluous" to electrodynamics and to optics.

Einstein found a solution to key problems in electrodynamics and optics only when he realized that an account of electromagnetic and optical action need not incorporate any assumptions about the mechanical properties of the ether.[xv] This was the assumption that required Hertz and Lorentz to include elements in their theories that, to Einstein, were unnecessary: Lorentz's still ether as absolute reference frame, and the vector quantities Hertz assigns to empty space.

*3.     Reconsidering experiments*

The question of what influence the Michelson, Michelson-Morley, and Fizeau experiments had on Einstein's 1905 paper depends on the answer to the question, How did Einstein interpret these experiments? Einstein would not (and indeed, could not) have used the Fizeau experiments to falsify Lorentz's and Hertz's ether theories: the experiments don't falsify those theories.

Instead, the brief remarks at the beginning of §6, that Einstein will neither appeal to the ether as stationary inertial frame, nor assign velocity vectors to empty space, indicate that Einstein used the lack of evidence for ether-matter interaction as delineating the outlines of, and substantive constraints on, the theory he should build. This is a substantive inference from the reconsideration of existing experiments. Hertz attempted to integrate the experiments on matter-ether interaction into his theory, but he ended up having to make a substantial idealizing assumption – assigning velocity vectors to empty space. Lorentz posited the seeming opposite, that the ether had no object-like properties at all, but was used as a rest frame of reference. Einstein's position in §6 of the 1905 paper is that neither of these theoretical strategies is necessary. But that is not to say that the problem that these theories were attempting to solve has an easy answer. It is to say the opposite – that the tradition of trying to use variables describing matter-ether interaction to construct explanations in electrodynamics and in optics had reached an impasse.

A major gain from re-examining experiments, or from examining them particularly closely in the first place, is to find mistaken or too-narrow inferences made by previous theories. These may indicate blind explanatory alleys or—in some cases, which can appear miraculous—new, unconsidered investigative paths. The examination of experiment can illuminate existing problems, but can also show that problems are lurking in the fit between existing theory and evidence, problems that have not yet been appreciated.

From an historical perspective, the more interesting the experiment, the more many-faceted are its consequences. An experiment can provide hints for the structure of a new theory, not just evidential support for that theory.[xvi] These hints can function in several ways:

1. They can give heuristic guidance as to the possible structure of the final theory;
2. They can help the scientist to avoid blind alleys of previous or merely possible theories;
3. They can provide the occasion for the scientist to see unexplored theoretical alternatives, to ask new questions, and to pose new problems—even when these alternatives, questions, and problems may not be related directly to the initial goal of the experiment.

Re-evaluating the evidential support for a theory is not the only reason to reconsider experiments. A new problem, a problem that existing theory does not consider or address, might be found in an existing experimental tradition. While Einstein was particularly gifted at evaluating scientific problems, reconsidering experiments and the problems they pose is an activity that might profitably be engaged in by historians and philosophers of science as well.

---

[i] Stachel 2002, 157-170, and Earman, Glymour and Rynasciewicz 1982, among many others, remark on this dearth of resources.

[ii] For an illuminating discussion of what Einstein knew of ether experiments, and how they influenced his thinking in the years leading up to 1905, see Stachel 2002, 171ff.

[iii] In this sense, I agree with the conclusions of Holton 1969.

[iv] See, e.g., Holton 1969, 136; Michelson 1927.

[v] Fresnel was also responding to Arago's demonstration from 1810 that the "earth's motion has no influence on the refraction of starlight in a prism" (see, e.g., Fizeau 1859, 350).

[vi] Mascart gives a brief description of the experiment here.

[vii] See Stachel 2005, 10.

[viii] See Darrigol 1999 and Darrigol 1993, 235ff.

[ix] "The medium is therefore capable of receiving and storing up two kinds of energy, namely, the 'actual' energy depending on the motions of its parts, and 'potential' energy, consisting of the work which the medium will do in recovering from displacement in virtue of its elasticity. The propagation of undulations consists in the continual transformation of one of these forms of energy into the other alternately" (Maxwell 1865, 463).

[x] Even if Einstein was not aware of the Michelson experiments, Hertz had reason to know about them in intimate detail. Hermann von Helmholtz had been Hertz's doctoral supervisor. Hertz finished his dissertation in 1880, then remained in Helmholtz's lab as a post-doctoral researcher for several years (Mulligan 2001, 147n.). In 1880, Michelson traveled to Berlin to work with Helmholtz. Michelson invented the interferometer and made his first experiments with it in Helmholtz's laboratory in Berlin in 1880-1881, while Hertz was working in the same laboratory. By 1890, Hertz had known about Michelson's early negative results for almost a decade.

[xi] See also a letter of September 3, 1889, from Hertz to Heaviside, cited by Mulligan (2001): "The motion of the ether relative to matter – this is indeed a great mystery. I thought about it often but did not get an inch in advance. I hope for experimental help; all that has been done till now has given negative results" (p. 147).

[xii] See also Mulligan 2001.

[xiii] See Stachel (2002), 157ff. and 171ff. for discussion of these points.

[xiv] See Hon and Goldstein 2005, 439, and 2005, 498ff, for a similar argument regarding Einstein's interpretation of an earlier paper of Hertz's.

[xv] The timeline of Einstein's reasoning, as reconstructed by Stachel 2002 and Earman, Rynasiewicz, and Glymour 1982, supports this reading.

[xvi] I would like to thank an anonymous reviewer for this journal for suggesting this way of putting it.